\documentstyle[preprint,tighten,eqsecnum,aps]{revtex}
\begin{document}
\draft
\preprint{UTS-DFT-94-09}
\title{Postmodern String Theory: Stochastic Formulation}

\author{A.Aurilia}
\address{
Department of Physics, California State Polytechnic University,\\ Pomona,
CA 91768}

\author{E.Spallucci}
\address{
Dipartimento di Fisica Teorica Universit\`a di Trieste,\\ INFN Sezione
di Trieste,\\  Strada Costiera 11, 34014 Trieste, Italy}

\author{I.Vanzetta}
\address{
Dipartimento di Fisica Teorica Universit\`a di Trieste,
\\  Strada Costiera 11, 34014 Trieste, Italy}
\maketitle

\begin{abstract}
In this paper we study the dynamics of a statistical ensemble of strings,
building on a recently proposed gauge theory of the string geodesic field.
We show that this stochastic approach is equivalent to the Carath\'eodory
formulation of the
Nambu-Goto action, supplemented by an averaging procedure over the family
of classical string world-sheets which are solutions of the equation of
motion. In this new framework, the string geodesic field is reinterpreted as
the Gibbs current density associated with the string statistical ensemble.
Next, we show that the classical field equations derived
from the string gauge action, can be obtained as the semi-classical limit of
the string functional wave equation. For closed strings, the wave equation
itself is completely analogous to the Wheeler-DeWitt equation used in
quantum cosmology. Thus, in the string case, the wave function has support
on the space of all possible spatial loop configurations. Finally, we show
that the string distribution induces a multi-phase, or {\it cellular}
structure on the spacetime manifold characterized by domains with a purely
Riemannian
geometry separated by domain walls over which there exists a
predominantly Weyl
geometry.
\end{abstract}
\bigskip
\pacs{PACS number: 11.17}


\section{Introduction and Synopsis}

The term {\it postmodern}, frequently encountered in art history, has been
lately applied to quantum mechanics to describe the renewed interest in
early semiclassical approximation schemes. In the words of Heller and
Tomsovic \cite{ht}, {\it  `` ... after the ``~modern~'' innovations have been
assimilated, the threads of premodern thought are always reconsidered.
Much of value may be rediscovered and put to new use. The modern context
casts new light on premodern thought, which in turn shades perspectives on
modernism."}

Those words aptly describe the spirit of this paper which has its roots in
the premodern string theory of Nambu \cite{y}, and Nielsen and Olesen
\cite{no}. That theory
led to the first successful description of hadrons as extended objects made
up of constituent quarks. Since then QCD has established itself as the local
gauge theory of quarks and gluons even though a generally accepted
quantitative account of the low energy binding process of quarks into
hadrons is still lacking. Over the years, the emphasis has shifted to
superstrings because of their alluring, but unfulfilled promise of
superunification of the fundamental forces. In more recent years, the
focus of attention has been on {\it cosmic strings}. Here again, the
tantalizing prospect is nothing short of explaining the large scale structure
of the universe in the framework of the inflation-driven Big Bang theory.
Even for cosmic strings there is a premodern theory, or {\it old picture},
based essentially on simulation codes. Admittedly, such numerical methods
provide thus far the most detailed knowledge about the dynamics of
a string network. However, their reliability, let alone their theoretical
value, may seem somewhat questionable. Indeed, even some of the original
investigators of cosmic strings have advocated the use of a more analytical
approach which might corroborate the numerical results \cite{kibble}.

Against this background, the aim of this paper is twofold: a) to construct a
new theory of relativistic strings building on the insights provided by the
premodern approach but without its shortcomings, and b) to investigate the
effect of the theory on the geometry of spacetime.

Central to our approach
is the premise that the dynamics of a string network, cosmic or otherwise,
is a random process at the outset \cite{emb}. From this statistical basis we
attempt to
construct a consistent theory of relativistic strings in terms of gauge
fields and quantum loop variables. The long range objective of this research
is perhaps best described as the formulation of a {\it stochastic quantum
field theory of relativistic strings}, which we regard as the paradigm of a
general approach to the dynamics of extended systems with any number of
dimensions.
The main points of the present approach can be summarized as follows :

1) Our string theory is initially defined in Minkowski space. However, we
will show that, in general, the geometry of the spacetime manifold is
shaped by the string distribution into a {\it domain}, or {\it cellular}
structure such that domain walls with a Weyl type geometry enclose {\it
voids}, or ``~pockets~''
of spacetime in a Riemannian phase.

2) In the quantum domain, our stochastic formulation does for strings what
the Wheeler-DeWitt equation does for quantum cosmology. In the string
case, loop configurations, interpreted as spatial sections of world -sheets,
correspond to the spatial 3-geometries in the Wheeler-DeWitt formulation
of quantum cosmology. A point-splitting regularization of the string wave
equation enables us to recover the classical field equations and to relate
the {\it quantum} probability amplitude defined in loop-space to the {\it
classical} string distribution function. In this technical sense we speak of
{\it stochastic} semi-classical formulation of string theory. This leads us
to the next and, perhaps, most fundamental point.

3) Stochastic formulations of quantum mechanics are just as old as
conventional quantum mechanics. This paper extends to relativistic strings
the stochastic approach used for point particles \cite{san}. On the technical
side, the
novelty of our approach stems from the use of Carath\'eodory's formulation
of
the Hamilton-Jacobi variational principle. As our analysis shows, even
though the Hamilton-Jacobi principle is deeply rooted in the deterministic
approach to classical mechanics, in Carath\'eodory's interpretation it takes
on a new meaning which opens the way to the stochastic formulation
suggested here. Since this point is the crux of our arguments, it seems
useful to expand on it at this introductory stage.
As is well known, one way to bridge the gap between the classical and
quantum domain is via the correspondence between wave phenomena and
geometric ray paths, and in this connection the importance of the
Hamilton--Jacobi
formulation of classical mechanics cannot be overemphasized. For our later
purposes, it is enough to recall the following two properties, i) the
Hamilton--Jacobi  equations can be derived in the WKB limit
from the Schr\"odinger equation, and, conversely, ii) the classical
Hamilton--Jacobi
wavefronts associated with the motion of a mechanical system represent
the surfaces of constant phase of the corresponding wavefunction.

In two previous papers, upon reexamination of premodern string
theory
\cite{y},\cite{no},\cite{egu1},\cite{n1},\cite{n2},\cite{Nambu},\cite{n4},
\cite{yh},
we have proposed a gauge field theory which admits extended
objects as singular solutions of the field equations \cite{noi},\cite{noi2}.
The theory is formulated in terms of two independent field variables which,
in the simplest case of a closed string, are a vector gauge potential
$A_\mu(x)$,
and an antisymmetric tensor $W^{\mu\nu}(x)$. Both fields  acquire a well
defined geometrical and physical meaning in the Hamilton--Jacobi\
formulation of string dynamics: once the field equations are
solved, $A_\mu(x)$ acquires the meaning of vector potential associated
with a family of extremal world-sheets, while
$W^{\mu\nu}(x)$ plays the role of string current. The field action, once
evaluated on such a solution, reproduces the Nambu-Goto action,
thus connecting the gauge formulation to the usual geometrical description
of the string motion.

However, from the postmodern vantage point,
the fundamental action is the gauge one, while the Nambu-Goto geometric
functional plays the role of {\it effective action} for the low-energy
string dynamics. In other words, postmodern string theory is {\it on-shell}
equivalent to the Nambu-Goto string model.

One of the objectives of this paper is to extend off shell the above on-shell
equivalence between the two formulations. Furthermore, we wish to show
that the
classical field equations for
$A_\mu(x)$ and $W^{\mu\nu}(x)$ can be derived as the WKB limit,
$\hbar\rightarrow 0$, of the functional wave equation for a closed,
bosonic string.

In this paper, both the off-shell equivalence between the geometric and
gauge string action,
and the semi-classical limit of the quantum wave equation are discussed in
the framework of the Carath\'eodory formulation of
the Hamilton--Jacobi action principle. A novel feature of our approach is an
averaging
procedure over the family of extremal world-sheets which are solutions of
the Hamilton--Jacobi equations. This procedure introduces,
at the classical level, a
random element in the string evolution. Each member of a family of
extremal
world-sheets is labelled
by a pair of parameters. We assume that the values of such parameters are
not equally likely but are statistically distributed.
In other words, rather than dealing with a single string, we consider a
statistical ensemble of them, not unlike a relativistic
fluid consisting of filamentary constituents.
The probability distribution, or fluid energy density, is an assigned
function at the classical level. {\it One of the main results of our analysis
is to relate such a classical function to the quantum probability amplitude,
hence connecting the statistical classical
behavior of the system described by our gauge
lagrangian, to the string quantum dynamics.} In view of this result, it is
tempting to interpret
the string fluid as a classical model of some quantum ground state, or
string condensate, leading us to a hydrodynamic picture of the string fluid
evolution in which $W^{\mu\nu}$ is interpreted as a divergence-free fluid
current, while $A^{\mu}$ plays the role of velocity potential.

The paper is organized as follows.

Sect. II is a pedagogical review of the Carath\'eodoryformulation of the
Hamilton--Jacobi\ action principle for a relativistic point particle.
It has the double purpose to introduce the relevant notation,
and to ``~translate~''
some of the mathematical terminology in a language which
is more familiar to physicists.

In Sect. III we discuss the Carath\'eodoryformulation
of string classical dynamics in {\it Minkowski space}, and introduce the
average over the family
of classical extremal surfaces. The final result is that we recover the
string gauge action from the Nambu-Goto area functional without the use of
classical solutions.

In Sect.IV we discuss the
semi-classical limit of the string functional wave equation. A point-
splitting regularization is introduced which enables us to recover
the classical field equations in the limit
$\hbar\rightarrow 0$. The distribution
function of classical extremals in parameter space is related to the square
of the string wave functional, and the quantum counterpart
of the classical current density is identified.

In Sect.V we show
that the string distribution induces a multi-phase, or {\it cellular},
structure on the spacetime manifold. In this structure, domains of
spacetime characterized by a Riemannian
geometric phase and a nearly uniform string distribution, appear to be
separated by domain walls. On these walls there exists a predominantly
Weyl geometry and across them the string density changes appreciably.

Throughout the paper we shall assume metric signature $-+++$, and
express physical quantities in  natural units. We use $\hbar\ne 1$
only in the semi-classical quantum formulae.

\section{Hamilton-Jacobi theory of the relativistic point-particle}

\subsection{Families of extremals and wave fronts}

The basic theoretical framework underlying our gauge theory of strings
 is the Carath\'eodory formulation of the Hamilton-Jacobi least action
principle.
The main features of that approach to classical dynamics can be
summarized as follows:
\begin{itemize}
\item
the variational principle takes into account a whole {\it family of
extremals} rather than a single classical solution
of the equations of motion;
\item
there is a remarkable duality between extremals and wave-fronts, in the
sense that the classical motion of the system can be equivalently described
either in terms of extremal trajectories, or in terms of propagation of the
associated wave-fronts;
\item
the essential dynamical variable is the {\it slope field}. This field
 assigns a tangent plane to each point of the classical
extremal;
\item
the whole theory can be elegantly formulated in terms of differential
geometry.
\end{itemize}
The primary purpose of this section is to introduce the basic idea of the
gauge model of strings and to set up the notation. We shall do so by
discussing first the case of a relativistic point particle. This case probably
represents the simplest example of a reparametrization invariant system
which is rich
enough to illustrate the effectiveness and simplicity of the method.

Except for a proportionality constant, the canonical action for a
relativistic object is the proper area of the world-manifold
spanned by the object during its evolution. Thus, the action corresponding to
  any extremal timelike world-hypersurface with a given initial and final
configuration is nonzero. That is true for a point--particle as
well.

The world--line swept by a pointlike object of rest mass $m$ is
determined by demanding that the functional
\begin{equation}
I[\gamma]=\int_{0,\gamma}^T d\lambda\,L(x,{\dot x};\lambda)\ ,\quad
L(x,{\dot x};\lambda)=-m\sqrt{-{\dot x}^\mu {\dot x}_\mu}\ ,\quad
{\dot x}^\mu={dx^\mu\over d\lambda}
\label{unouno}
\end{equation}
 be stationary under small variations of the path $\gamma$ connecting
the initial and final points
$x^\mu_0=x^\mu(0)$ and $x^\mu=x^\mu(T)$.

The extremal trajectory is a solution of the Euler-Lagrange equation
\begin{equation}
{dP_\mu\over d\lambda}=0\ ,
\label{unodue}
\end{equation}
where the particle four-momentum is defined as follows
\begin{equation}
P_\mu\equiv{\partial L\over\partial {\dot x}^\mu}=
m{{\dot x}_\mu\over\sqrt{-{\dot x}^2}}\ ,
\label{unotre}
\end{equation}
and, because of reparametrization invariance, is subject to the
constraint
\begin{equation}
-P_\mu P^\mu=m^2\ .
\label{unoquattro}
\end{equation}
Equations (\ref{unodue}),(\ref{unoquattro}) can be immediately solved,
and give:
\begin{equation}
P_\mu=m{v_\mu\over\sqrt{-v^2}}\ ,
\label{unocinque}
\end{equation}
where $v_\mu$ is a constant vector field. The classical solution
(\ref{unocinque}) of the Lagrange equations (\ref{unodue}), under
the constraint (\ref{unoquattro}), introduces $v^\mu$ as the
{\it slope field}, so-called because once eq.(\ref{unocinque}) is re-inserted
into equation (\ref{unotre}), it matches the tangent vector along the
particle
world--line. Indeed,
once the boundary conditions are assigned, and the
slope field is obtained from the Lagrange equations,
one obtains through equation (\ref{unotre}) an extremal curve solving the
given
dynamical problem.
In the free particle case the slope field
is constant and the world-line is  a geodesic
\begin{eqnarray}
&&x^\mu(\lambda)={(x-x_0)^\mu\over T}\lambda+x^\mu_0\ ,\nonumber\\
&&P_\mu=m{(x-x_0)_\mu\over\sqrt{-(x-x_0)^2}}\ .
\label{unosei}
\end{eqnarray}
Accordingly, the tangent vector field is constant and so is
${\dot x}^\mu(\lambda) =(x-x_0)^\mu/T$. But this is only because the
particle is free.
In the more general case in which external forces are present, both the
four-velocity and momentum are restricted to vary along the extremal, i.e.
${\dot x}^\mu={\dot x}^\mu(\lambda)$, ${\dot x}_\mu=P_\mu(\lambda)$.
However, the Hamilton-Jacobi approach enables one to
extend the four momentum to a {\it four} dimensional sub-manifold
of Minkowski space $M^{4)}$ by taking into account that eqs.(\ref{unosei})
describe a
{\it family} of classical solutions parametrized by the initial particle
position
and momentum. For later convenience, we are interested in the initial
spatial position of the particle defined by the
three-vector $\vec x_0$. Then, any point $x^\mu$ belonging to a classical
extremal curve, will become a function of four new ``~coordinates~''
\begin{equation}
x^\mu=x^\mu(y)\ ,\qquad y^\alpha\equiv (\lambda,\vec x_0)\ .
\label{unosette}
\end{equation}
Notice that varying $\lambda$ amounts to move $x$ along one definite
extremal, while variation of $\vec x_0$ corresponds to select a different
path within the family of classical extremals. In this respect
$\partial/\partial x_0^k$ represent ``~orthogonal directions~'' with respect
to $\partial/\partial\lambda$. If the mapping between $x^\mu$ and
$y^\alpha$
is one-to-one, then the classical solution (\ref{unosette}) represents an
invertible change of coordinates. Hence, when a family of classical
solutions is properly taken into account, it is possible to define a
corresponding momentum field $P_\mu(x)$, or velocity field $v^\mu(x)$,
having support over a four dimensional spacetime region.
The Carath\'eodory approach to the calculus of variations provides a well
defined
procedure to construct these ``~extended~'' fields, and is free of
integrability problems.

One of the most interesting features of the Hamilton-Jacobi theory is to
associate {\it wave fronts} to pointlike objects.
To introduce these new geometrical objects
we recall that the Jacobi action, or dynamical phase, is obtained
by evaluating the line integral (\ref{unouno}) along the classical solution
(\ref{unosei}).
\begin{equation}
I(x;x_0)=\int_0^T d\lambda\,L(x(\lambda),{\dot x}(\lambda))= -m\sqrt{-(x-
x_0)^2}
\equiv S(x)\ ,
\label{unootto}
\end{equation}
Hence $S$ defined above, becomes an ordinary function of the geodesic end-
points: $x$ which is to be understood as a variable and $x_0$ which should
be
interpreted as a parameter fixed by the initial conditions.  Comparing
eqs.(\ref{unosei}) and (\ref{unootto}), one sees that $S(x)$
must satisfy the relativistic Hamilton-Jacobi equations
\begin{eqnarray}
&&\partial_\mu S\partial^\mu S=-m^2\ ,\label{onde}\\
&&{\dot x^\mu\over\sqrt{-\dot x^2}}=
{1\over m}\partial^\mu S(x)\ ,
\label{unotto}
\end{eqnarray}
where, $\partial_\mu=\partial/\partial x^\mu$.  The Hamilton-Jacobi wave
fronts are now defined as the
three-surfaces of constant dynamical phase, i.e.
\begin{equation}
S(x)=\,const.
\label{unonove}
\end{equation}
The surfaces (\ref{unonove}) obey the wave type evolution equation
(\ref{onde}), and are transverse with respect to the congruence of
spacetime curves solving eq.(\ref{unotto}).
In this sense a given extremal can be ``~embedded~'' into a system
of wave fronts, any point on a constant phase surface being the intersection
of one, and only one,  particle world line. Then, once the family
of Hamilton--Jacobi\ wave fronts is known, the gradient of $S(x; x_0)$
determines
a slope field in the local sense, i.e. defined at each point over
the surfaces of the family. Such a slope field determines a family of
particle world-lines, crossing the Hamilton--Jacobi\ wavefronts.
The geometric object formed by the family of extremals and
the associated wave fronts constitutes what is called the {\it complete
Carath\'eodoryfigure}, and provides a unique characterization of
the solution in Carath\'eodory's own version of the
Hamilton-Jacobi variational principle.

 It is worth emphasizing, parenthetically, that equation (\ref{onde}) was
obtained after solving the equation of motion,
which in the case of a point particle is very simple. However, in order to
deal with the string case we need an algorithm that enables us to
recover  equation (\ref{onde}) without knowing the explicit form of
the solution. This can be done by considering the hamiltonian form of the
action integral:
\begin{equation}
I[\gamma]=\int_0^T d\lambda \left[\,P_\mu\,\dot x^\mu-
N(\lambda)(P^2+m^2)
\right]
\ ,\label{pam}
\end{equation}
where, $N(\lambda)$ is a lagrange multiplier enforcing the constraint
(\ref{unoquattro}). If we evaluate $S(x)$ on a classical trajectory,
say $x=y(\lambda)$, then we find
\begin{equation}
I(x,x_0)=\int_{x_0}^x P_\mu(y)\,dy^\mu=
\int_0^T d\lambda\, L\left(y(\lambda),\dot y(\lambda)\right)\ .
\label{scl}
\end{equation}
Since $S(x)$ is a function of the extremal
end-points only, then, when the path is varied, the requirement that
$I(x,x_0)$ be
an extremum, leads to:
\begin{equation}
\int_0^T d\lambda\,\partial_{[\,\mu}P_{\nu]}\,
\dot y^\nu\,\delta y^\mu+
\left[P_\mu\,\delta y^\mu\right]^x_{x_0}=0 \ .
\end{equation}
Here, the variation of the trajectory end-points is independent of the path
deformation. Then, setting
$\delta y^\mu(T)=dx^\mu$, and
$\delta y^\mu(0)=0$,  we obtain
\begin{eqnarray}
&&\partial_{[\,\mu}P_{\nu]}{ dy^\nu\over d\lambda}=0\ ,\label{hj1}\\
&&{\partial S\over \partial x^\mu}=P_\mu(x)\ .\label{hj2}
\end{eqnarray}
The above set of equations tells us that along a classical trajectory
the linear momentum can be expressed as the gradient of a scalar function.
This is exactly what we did in equations (\ref{unotto}, \ref{onde}).

This result, simple as it is in this case,
can be translated in the language of differential geometry as follows:
the momentum one-form $\theta(x)\equiv P_\mu(x)dx^\mu$ is closed along
a
classical trajectory so that, at least locally, it is an exact one-form
\begin{equation}
d\theta\left(x=x(\lambda)\right)=0\,\rightarrow
\theta\left(x=x(\lambda)\right)=
dS\left(x=x(\lambda)\right)=L\left(x(\lambda),\dot x(\lambda)\right)
d\lambda\ .\label{esatta}
\end{equation}
For reparametrization-invariant theories,
the last equality
can be written as:
\begin{equation}
\theta(x=x(\lambda))=\dot x^\mu(\lambda)\partial_\mu S(x) d\lambda\ .
\label{lag}
\end{equation}
In this form, the above results apply to the string case as well.


\subsection{The Carath\'eodoryleast action principle}

The ideas expressed above in terms of the
one-form $\theta(x)$, find
an immediate application
in the Carath\'eodory formulation of the Hamilton-Jacobi theory.
The main
point is to modify the lagrangian according to:
\begin{equation}
L(\dot{x})\longrightarrow L^*({\dot x})=L(\dot{x})-\frac{dS(x)}{d\lambda}=
L(\dot{x})-\dot{x}^\mu\partial_\mu S(x)\label{newlag}.
\end{equation}
As it is well known, the addition of a
total derivative does not affect the equations of
motion and therefore $L$ and $L^*$ are dynamically equivalent. The principle
of
least action {\it applied to the new lagrangian} yields:
\begin{eqnarray}
0=\delta I^*(x,x_0)&=&\int_0^T
d\lambda\, L^*(\dot{x})=\nonumber\\
&=&\int_0^T
d\lambda\,\delta x^\mu(\lambda)\left[
{\partial L\over\partial x^\mu}-{d\over d\lambda}
{\partial L\over\partial\dot x^\mu}\right]+\nonumber\\
&+&\left[\delta x^\mu(\lambda)\left({\partial L\over\partial\dot x^\mu}-
\partial_\mu S(x)\right)\right]_0^T\ .
\label{bracket}
\end{eqnarray}
As a matter of fact, even if {\em the variation is restricted to a family of
solutions of the
Lagrange equations}, with the boundary condition
$\delta x^\mu(0)=0$, we still have to impose
\begin{eqnarray}
\delta I^*\Big\vert_{x=x(\lambda)}=0
&\rightarrow& P_\mu(x)
=\partial_\mu S(x)\ .
\label{jac}
\end{eqnarray}
 i.e., the requirement that the momentum is the gradient
of the Jacobi phase.
{}From the definitions (\ref{newlag}), (\ref{jac}), it follows that the
solutions
of the above variational problem can be obtained   from the condition
\begin{equation}
\left({\partial L^*\over\partial {\dot x}^\mu}\right)_{x=x(\lambda)}=0\ .
\label{tre}
\end{equation}
 The proper length lagrangian is invariant under reparametrization,
therefore
it is homogeneous of degree one with respect to the generalized velocity
${\dot x}^\mu$, so that
\begin{equation}
L^*(x(\lambda),{\dot x}(\lambda))=0
\label{trebi}
\end{equation}
 follows from equation (\ref{tre}). For a more general non-
reparametrization
invariant theory,
eqs.(\ref{tre}) and (\ref{trebi}) represent the {\it fundamental equations
of the calculus of variation,} the associated complete figure consisting of
a three-parameters congruence of extremals intersecting a one-parameter
family of wave-fronts. The usefulness of this modified variational
approach and its connection with field theory, is appreciated when one
considers not a single particle but a {\it statistical ensemble} of them.

\section{Carath\'eodory string action}

\subsection{Hamilton-Jacobi theory for the closed string}

A closed string is formally described by a two dimensional domain $D$ in
the space
of the parameters ${\xi}^a=(\tau,\sigma)$ which represent local coordinates
on the string world-sheet, and by an embedding $\Omega$ of $D$ in
Minkowski
spacetime $M$, that is
$\displaystyle{\Omega: {\xi}\in D\rightarrow \Omega({\xi})=X^\mu({\xi})\in M}$.
We assume that the domain $D$ is simply connected and bounded
by a {\it spacelike} curve
$\displaystyle{\gamma:{\xi}^a={\xi}^a(s)}$. Furthermore, we can choose the
parameter
$s$ to vary
in the range $0\le s\le 1$. Then, $\Omega$ maps the boundary $\gamma$
into
the {\it spacelike loop} $C$ in spacetime:
$\displaystyle{x^\mu=x^\mu\left({\xi}(s)\right)=x^\mu(s)\ ;\quad
x^\mu(s+1)=x^\mu(s)}$.
The resulting picture describes a spacetime world-sheet whose
only ``~free end-point~'' is the spacelike loop $x=x(s)$.
The Euler-Lagrange equations, derived from
\begin{eqnarray}
L(x,\dot x;{\xi})&=&
-m^2\sqrt{-\gamma({\xi})}\ ,\quad
\gamma({\xi})={1\over 2}\dot x^{\mu\nu}\dot x_{\mu\nu}\ ,\nonumber\\
\quad \dot x^{\mu\nu}&=&\epsilon^{ab}\partial_a x^\mu\partial_b x^\nu\ ,
\qquad m^2={1\over 2\pi\alpha'}\ ,
\label{ng}
\end{eqnarray}
read
\begin{equation}
\epsilon^{ab}\partial_a \Pi_{\mu\nu}\partial_b x^\nu=0\ ,\qquad
\Pi_{\mu\nu}\equiv {\partial L\over\partial\dot x^{\mu\nu}}=
m^2{\dot x_{\mu\nu}\over\sqrt{-\gamma({\xi})}}\ ,
\label{amom}
\end{equation}
and, the {\it area momentum} $\Pi_{\mu\nu}$, canonically conjugate
to $\dot x^{\mu\nu}$, satisfy the ``mass-shell'' constraint
\begin{equation}
-{1\over 2}\Pi_{\mu\nu}\Pi^{\mu\nu}=m^4\
\label{mass}
\end{equation}
because of the reparametrization invariance of the theory.
The two equations (\ref{amom}, \ref{mass}) define the string slope field
$\Phi_{\mu\nu}({\xi})$ as
\begin{equation}
\Pi_{\mu\nu}({\xi})=m^2{\Phi_{\mu\nu}({\xi})\over\sqrt{-|\Phi|^2}}\ ,\qquad
|\Phi|^2\equiv {1\over 2}\Phi_{\mu\nu}({\xi})\Phi^{\mu\nu}({\xi})
\label{slope}
\end{equation}
where,
\begin{equation}
\epsilon^{ab}\partial_a {\Phi_{\mu\nu}({\xi})\over\sqrt{-|\Phi|^2}}
\partial_b x^\nu=0\ .
\end{equation}
Dealing with a spatially extended object means that the slope field
is now a bi-vector field, which assigns to every point a plane.
Once such a field of planes is known, the extremal surface swept by the
string evolution is given by
the equation
\begin{equation}
\epsilon^{ab}\partial_a x^\mu\partial_b x^\nu=\Phi^{\mu\nu}({\xi})\ ,
\label{estr}
\end{equation}
which represents the condition that, at each point, the world-sheet tangent
element overlaps with the corresponding tangent plane. Equation (\ref{estr})
suggests
a hydrodynamic picture of the string as a vortex line co-moving in
a viscous fluid described by the velocity field $\Phi^{\mu\nu}({\xi})$.
In this picture, the matching equation (\ref{estr}) represents the
no--slipping condition for the string within the fluid.

To obtain the Hamilton--Jacobi\ equations for the string, we turn to the
hamiltonian
form of the Nambu-Goto action
\begin{equation}
I[X({\xi})]={1\over 2}\int_{X({\xi})} dx^\mu\wedge dx^\nu \Pi_{\mu\nu}(x)-
\int_D d^2{\xi}\, N({\xi})\left[{1\over 2}\Pi_{\mu\nu}\Pi^{\mu\nu}+m^4
\right]\ ,
\label{hng}
\end{equation}
and consider the variation of $I$ among the classical solutions of equation
(\ref{estr}). Thus, $I$ becomes a functional of the boundary $C$ alone,
the constraint (\ref{mass}) is satisfied, and we find
\begin{equation}
\delta I[X({\xi})]\rightarrow
\delta I[C]=-\int_D d^2{\xi}\partial_{[\,\lambda}\Pi_{\mu\nu]}
\dot x^{\lambda\mu}\delta x^\nu +\oint_C \Pi_{\mu\nu}dx^\mu\delta x^\nu\
{}.
\label{bs}
\end{equation}
The first term in (\ref{bs}) must vanish, otherwise a dependence on the
whole string world-sheet, and not on its boundary alone, would be implied.
The second term represents the response of the action to a variation of
the boundary. Then,
\begin{eqnarray}
&&\partial_{[\,\lambda}\Pi_{\mu\nu]}\,\dot x^{\lambda\mu}=0\
,\label{bianchi}\\
&&P_\mu(s)\equiv
{1\over\sqrt{x^{\prime\,2}}}{\delta I[C]\over\delta
x^\mu(s)}=\Pi_{\mu\nu}(x(s))
{x^{\prime\,\nu}\over \sqrt{x^{\prime\,2}}}\ .
\label{mom}
\end{eqnarray}
Equation (\ref{bianchi}) is nothing but the Bianchi identity for
$\Pi_{\mu\nu}$
projected over an extremal surface, and tells us that strings are somehow
related to gauge potentials. {\it Here is where one encounters the problem
of ``~thickening~'' the world sheet}\cite{stach}: it would be tempting to
conclude that  (\ref{bianchi})
is satisfied not only  on the string world-sheet but over a {\it spacetime
region}, in which case one would define,
$\Pi_{\mu\nu}(x)=\partial_{\,[\mu}A_{\nu]}(x)$. However, this conclusion is
too hasty because for a two-surface, ${\rm det}\,\dot x^{\mu\nu}=0$
\cite{noi}, and (\ref{bianchi}) admits solutions for which
$\partial_{[\,\lambda}\Pi_{\mu\nu]}\ne 0$ as well. The vanishing of
${\rm det}\dot x^{\mu\nu}$ is due to the low dimensionality of the string
world-surface: in two dimensions there is no totally anti-symmetric three
index tensor. To circumvent this difficulty, we have to take into account a
two-parameter congruence of surfaces,
$x^\mu=x^\mu(\tau,\sigma,\theta_1,\theta_2)$,
rather than a single extremal world-sheet, and treat the labels
 $\theta_1,\theta_2$ as additional coordinates. Then, ${\rm det}\dot
x^{\mu\nu}$
becomes a function of four coordinates, instead of two, and is generally
different from zero. The congruence of extremal world-surfaces can be
interpreted
as a mapping from the extended parameter space to the physical spacetime.
Then, the two-form
\begin{equation}
\theta(x)={1\over 2}\Pi_{\mu\nu}(x)\, dx^\mu\wedge dx^\nu\ ,
\label{2pi}
\end{equation}
is closed over a two-parameter family of classical solutions
$x^\mu=x^\mu({\xi}^a,\theta^i)$, and, at least locally, can be written
in terms of a gauge potential $A_\mu(x)$
\begin{equation}
d\theta\left(x=x({\xi},\theta)\right)=0\ ,\rightarrow
\Pi_{\mu\nu}(x)=\partial_{[\,\mu}A_{\nu]}(x)\ .
\end{equation}
In other words, the action for a family of classical solutions, i.e.
\begin{equation}
I[C]={1\over 2}\int_{X({\xi})} dx^\mu\wedge dx^\nu\, \Pi_{\mu\nu}(x)\ ,
\end{equation}
is a boundary functional only if
$\Pi_{\mu\nu}$ can be written as the curl of a vector potential, in which
case
$I[C]$ becomes a loop integral
\begin{equation}
I[C]=\oint_C dx^\mu A_\mu=
\int_D\, d^2{\xi} L\left(x({\xi}),\dot x({\xi})\right)\ .
\end{equation}
Now, we still have the freedom to fix the rank\footnote{ We recall that
the rank {\it r} of a p-form $\omega$ is the minimal number {\it r} of
linearly
independent 1-forms in terms of which $\omega$ can be expressed.}
of $\theta(x)$.
In four dimensions a 2-form has either
rank-four or rank-two, which means that $\theta$ can be written
in terms of four or two independent one-forms. The rank-four option
means
\begin{equation}
\theta(x)=dS^1\wedge dT^1+dS^2\wedge dT^2  .
\label{r4}
\end{equation}
But, in this case the wavefronts degenerate into points,
so that the complete Carath\'eodoryfigure associated with the string
motion becomes meaningless. A first attempt to solve this problem was
proposed by
 Nambu \cite{Nambu}, who tried to generalize the
Hamilton--Jacobi\ formalism for non-relativistic system to the string case as
well.
The starting point is the non-reparametrization invariant Schild action
\cite{Schild} which leads to a six-dimensional Hamilton--Jacobi\ formalism
where
$S^{i)}=S^{i)}(x,{\xi})$, and $T^{i)}=T^{i)}(x,{\xi})$, $i=1,2$. However, this
approach only apparently solves the integrability problem. In fact, the
resulting wavefronts are two-dimensional surfaces, while the general
rule for mechanical systems of this kind would require the dimension
of the wavefronts to be four \cite{karin}.

The alternative possibility, that we consider, is to preserve
reparametrization invariance
and change the lowest rank for $\theta$.

By definition, a wavefront is the integral sub-manifold corresponding
to the characteristic subspace $C(\theta)$. A vector field $Y$ belongs
to $C(\theta)$ if it annihilates both $\theta$ and $d\theta$, i.e.
$\displaystyle{i(Y)\theta=0\ , i(Y)d\theta=0}$. According to the
general definition above, the wavefront will be $(D-c)$--dimensional
where D is the spacetime dimension and $c=dim\,C(\theta)$ is
the {\it class} of $\theta$\cite{ka}. If $\displaystyle{ d\theta=0\ ,
\rightarrow c=r=rank(\theta)}$, and the wavefronts are $(D-r)$--
dimensional
hypersurfaces. Hence, by choosing for $\theta$ the closed, rank--two,
two--form
\begin{eqnarray}
&&\theta(x)=dS\wedge dU\ ,\label{r2}\\
&&\theta\left(x=x({\xi})\right)=
\dot x^{\mu\nu}\partial_\mu S\partial_\nu U\, d^2{\xi}
\nonumber\\
&&\phantom{\theta\left(x=x({\xi})\right)}=
L\left(x({\xi}),\dot x({\xi})\right)\, d^2{\xi}
\end{eqnarray}
the class c$=$rank$=$2 and the
wave-fronts are $D-r=$two-dimensional, integral manifolds,
given as solution of the equations
\begin{eqnarray}
S(x)=const. \equiv \sigma_1\ ,\label{f1}\\
U(x)=const. \equiv \sigma_2\ .\label{f2}
\end{eqnarray}
The surfaces defined by (\ref{f1},\ref{f2}) satisfy the ``wave equation''
\begin{equation}
{1\over 2}\partial_{\,[\mu}S\partial_{\nu]}U
\partial^{\,[\mu}S\partial^{\nu]}U=-m^4\ ,
\label{hjs}
\end{equation}
and, together with the family of extremal surfaces solving
\begin{equation}
{\dot x^{\mu\nu}\over\sqrt{-\gamma({\xi})}}={1\over m^2}
\partial^{\,[\mu}S\partial^{\nu]}U\ ,
\end{equation}
form the complete Charath\`eodory figure associated with the
string motion.

Turning now to the boundary eq.(\ref{mom}), we note that it provides the
Jacobi definition of the string momentum
$P_\mu(s)$ as the dynamical variable canonically conjugate to the
world-sheet boundary deformation. By inverting equation (\ref{mom})
we can give meaning to the area momentum on the boundary
\begin{equation}
\Pi_{\mu\nu}(s)={P_{[\mu}(s)x'_{\nu]}(s)\over\sqrt{ x^{\prime\,2}}}\ ,
\label{pibordo}
\end{equation}
where we have taken into account the orthogonality relation
$P_\mu(s)x'{}^\mu=0$. Equation (\ref{pibordo}) shows that
the string momentum satisfies the mass-shell relation
\begin{equation}
\left(\int_0^1 ds\,\sqrt{x^{\prime\,2}}\right)^{-1}
\int_0^1 ds\,\sqrt{x^{\prime\,2}}\, P_\mu(s) P^\mu(s)=-m^4\ ,
\label{media}
\end{equation}
following from equation (\ref{mass}), and an average over the loop
parameter
$s$ which makes eq.(\ref{media}) reparametrization invariant.
The relation between classical momenta will
be a key ingredient in the quantum formulation of string dynamics
through the Correspondence Principle.

\subsection{ The averaged Charath\'eodory variational principle}

The Carath\'eodory formulation of the HJ variational principle is obtained
from the modified action functional:
\begin{eqnarray}
I[X({\xi})]&=&\int_D d^2{\xi}\,L(\dot X)-\oint_C S(x)\,dU(x)\ ,\nonumber\\
&=&\int_D d^2{\xi}\,\left[L(\dot X)- {1\over 2}{\dot x}^{\mu\nu}
\partial_{\,[\mu}S\partial_{\nu]}U\right]   \ ,\nonumber\\
&=&\int_D d^2{\xi}\,L^*(\dot X)\ .
\label{dueuno}
\end{eqnarray}
As far as the variational problem is concerned, the presence of
an additional boundary term is irrelevant, and the classical solutions
will be the same ones that follow from the  Nambu-Goto action. However,
the presence of the slope field in Carath\'eodory's formulation of the least
action principle, is instrumental in connecting the description
of strings dynamics in terms of mechanical variables with the formulation
in terms of a gauge potential. To appreciate this point which is essential in
the postmodern gauge interpretation of string theory, observe that
the slope field which minimizes the action (\ref{dueuno})
must be a solution of the equation
\begin{equation}
\frac{\partial L^*(x,{\dot x};{\xi})}
{\partial{\dot x}^{\mu\nu}}\Big\vert_{x=x({\xi})}=0
\ ,\rightarrow \Pi_{\mu\nu}({\xi})=\partial_{\,[\mu}S\partial_{\nu]}U
\ ,\rightarrow L^*(x,{\dot x};{\xi})\Big\vert_{x=x({\xi})}=0
\label{211}\ .
\end{equation}
Equation (\ref{211}) requires $S(x)$ and $U(x)$ to be solutions of
the Hamilton--Jacobi\ equation (\ref{hjs}) and the family of extremal surfaces
$x=x({\xi})$ to solve equation (\ref{hjs}). Thus, the condition (\ref{211})
defines the same complete Carath\'eodory figure associated with the string
motion
as the hamiltonian action (\ref{hng}), and is therefore equivalent to it. The
novel feature of the string action {\it a--la Carath\'eodory} is the presence
of an ``interaction term'' between the loop current $\displaystyle{j^\mu(x;C)=
\int_0^1ds\, x^{\prime\,\mu}\,\delta^4[x-x(s)]}$,
and the ``~external electromagnetic gauge potential~''
$\displaystyle{A_\mu=S\partial_\mu U(=-U\partial_\mu S+\partial_\mu(SU))}$.
With this interaction term the model becomes invariant under the gauge
transformation $\displaystyle{\delta A_\mu=\partial_\mu\Lambda}$, and,
furthermore, the corresponding ``~electromagnetic
strength~'' $F_{\mu\nu}$ has minimal {\it rank-two} \cite{ka}, which
is the distinctive feature of a string-like field excitation. All this suggests
that it should be possible to give a gauge form to the Nambu-Goto action as
well. This objective is achieved by considering a statistical ensemble of
strings, in other words, we switch from a single string evolution problem to
the dynamics of a ``Gibbs ensemble'' of strings. For such a
purpose we need:

i) a {\it dense} set ${\cal P}$
of spacetime surfaces $x^\mu=x^\mu({\xi})$, i.e., a family of extremals
covering at least a four-dimensional region $G^{4)}\subset M^{4)}$. We will
refer to
it as the ``path-space of the string''.
After imposing a particular variational principle, ${\cal P}$ will become a
subset
of all the possible world-sheets swept by the string evolution;

ii) an invariant measure given over this set assigning the
probability for the string to move along a given trajectory.
\smallskip
i) The path-space ${\cal P}$ can be constructed as follows.

Consider the above
mentioned family of non-intersecting surfaces: since the surfaces are
two-dimensional objects, and since they are required to form a dense set
covering a four-dimensional submanifold, it will consist of a {\it two
parameter}, say $u^1,u^2$, family of surfaces.
The rationale for introducing ${\cal P}$ is to set up a one--to--one
correspondence
between $(x^0,x^1,x^2,x^3)$ in $G^{4)}$ and $(\sigma,\tau;u^1,u^2)$ in
parameter space. Still, their ``form''
remains free, i.e. $G^{4)}$ can be covered by planes, or by spherical
shells or whatever.
As a characterization of the whole family, take, e.g., the tangent elements
$\dot{x}^{\mu\nu}({\xi})$ to every surface, and require them to satisfy:
\begin{equation}
m^2\frac{\dot{x}_{\mu\nu}({\xi})}{\sqrt{-\gamma({\xi})}}=
\partial_{\,[\mu}S\partial_{\nu]}U\ .
\end{equation}
For the moment, $S(x)$ and $U(x)$ are two {\it generic} functions:
they shall become the two Jacobi potentials of the H-J theory only after
certain conditions are met.
\smallskip
ii) As an invariant measure
(under reparametrization of the surfaces belonging to ${\cal P}$),
introduce a positive definite weight $\mu(u)$ defined on the
space of parameters $u^1,u^2$. Any pair $u^1,u^2$ will label
a single surface within ${\cal P}$. Then, we can define the probability
for a string to move along a sample path with parameters in the
subset $\{B\}$ in  parameter space, as:
\begin{equation}
prob(B)\equiv\int_{B}d^2u\ \mu(u)\ .
\end{equation}
The peaks of $\mu(u)$ correspond to more probable values of the
parameters, i.e., to statistically preferred classical paths. Thus,
even if the evolution remains strictly deterministic, a random
element is introduced into the model. At this stage the density $\mu(u)$
is an arbitrarily assigned positive definite function. However, in the last
part of the paper, we will show
how this distribution is fundamentally  related to the quantum dynamics of
the system.

Given the action for a two-parameter family of surfaces randomly
distributed
in the parameter space, the slope field solving the least action
principle is obtained by averaging the Carath\'eodory action
over the parameters $u^a$
\begin{equation}
\delta \langle I \rangle=\delta\int d^2u\ \mu(u)\int_D d^2{\xi} \,
L^*\left(\dot{x}({\xi};u)\right)=0\label{average}.
\end{equation}
Since $\mu(u)$ does not depend on ${\xi}^a$ and is positive definite,
eq. (\ref{average}) is just
$\int\,\hbox{eq.(\ref{dueuno})}\,\mu(u)\, d^2 u$, and gives the usual
equations
of motion upon variation of the appropriate variables. This means that,
imposing the
``~average minimum action principle~'', the family of surfaces in ${\cal P}$
becomes the congruence of extremals of the Carath\'eodory complete figure,
and $S(x)$, $U(x)$ become the two Jacobi phases.

The reason for integrating over $d^2u$ becomes clear as soon as
we  transform the variables $({\xi};u)$ to the $x^\mu$. Eq.(\ref{average})
then reads:
\begin{equation}
\langle I\rangle=\int_{G^{4)}}d^4x\,
\mu\left(u(x)\right)\,Z(x)\,  L^*\left({\dot x}^{\mu\nu}(x)\right)=
\int_{G^{4)}}d^4x\, L^*\left(\rho(x){\dot x}^{\mu\nu}(x)\right)\ ,
\end{equation}
where $\dot{x}^{\mu\nu}(x)\equiv \dot{x}^{\mu\nu}({\xi}(x),u(x))$, and
$Z(x)\equiv\partial({\xi}^1,{\xi}^2;u^1,u^2)/\partial (x^0,x^1,x^2,x^3)$.
Moreover, in view of the homogeneity of the lagrangian,
$\rho(x)\equiv Z(x)\mu\left(u(x)\right)$ has been taken into the argument
of $L^*$. Finally,
introducing the ``~Gibbs ensemble associated current~''
\begin{equation}
W^{\mu\nu}(x)\equiv \mu(x)Z(x){\dot x}^{\mu\nu}(x)\equiv \rho(x)
{\dot x}^{\mu\nu}(x)\ ,
\label{duedodici}
\end{equation}
{\it we can write $\langle I\rangle$ as the action for
a relativistic field theory:}
\begin{equation}
I_{field}=\langle I\rangle =\int d^4x\,\left[-m^2\sqrt{-{1\over 2}
W^{\mu\nu}(x)W_{\mu\nu}(x)}-{1\over 2}
W^{\mu\nu}(x)\partial_{[\,\mu}A_{\nu\,]}\right]\ .
\label{duetredici}
\end{equation}
This
action
was introduced by the authors \cite{noi}
as a gauge model with a {\it single string}
slope-field $W^{\mu\nu}$
proportional to the classical string current, i.e.
\begin{equation}
W^{\mu\nu}=const. \int_D d^2{\xi}\,\delta^{4}(x-x({\xi})){\dot
x}^{\mu\nu}({\xi})
\end{equation}
which represents a special solution of the classical field equations
\begin{equation}
m^2{W_{\mu\nu}(x)\over\sqrt{-{1\over
2}W^{\rho\sigma}(x)W_{\rho\sigma}(x)}}=
\partial_{\,[\mu} A_{\nu]}(x)\ ,\label{duequattordici}
\end{equation}
\begin{equation}
\partial_\mu W^{\mu\nu}(x)=0\ .
\label{duesedici}
\end{equation}
However, within the stochastic approach advocated here, a more general
solution can be given.
Indeed, with the definition
$\overline{\rho}(x)\equiv m^2\,\sqrt{-\gamma}\,\rho(x)$,
it is immediate to check that
$\displaystyle{W^{\mu\nu}(x)=(\overline{\rho}(x)/m^4) F^{\mu\nu}(x)}$
is a solution of eq.
(\ref{duequattordici}), (\ref{duesedici}) if $F^{\mu\nu}(x)$ satisfies
the H--J equation
\begin{equation}
-{1\over 2}F_{\mu\nu}(x)F^{\mu\nu}(x)+m^4=0\ ,\label{duesedicibis}
\end{equation}
and  the ``~current~'' $\overline{\rho}(x)\,F^{\mu\nu}$ is divergenceless
\begin{equation}
\partial_\mu\left(\overline{\rho}(x)\,F^{\mu\nu}(x)\right)=0\ .
\label{duediciasette}
\end{equation}

{\it Rather than the slope-field associated to the single string case,
$W^{\mu\nu}(x)$ represents, now, the divergence--free
Gibbs current for a statistical system of filamentary structures.}
One can visualize this system as a {\it fluid of relativistic strings}
covering the spacetime manifold with a distribution density
$\overline{\rho}(x)$ and a
velocity field $\sim F^{\mu\nu}$. In this {\it hydrodynamic} interpretation
of
the action (\ref{duetredici}), two very general questions come immediately
to mind. The first pertains to the classical domain, the second to the
quantum domain: if Einstein's theory of gravitation is any guide, one would
expect that the geometry of spacetime is shaped by the matter--string
distribution. We would like to determine how. In other words, is the
classical spacetime manifold purely Riemannian?, {\it and} is there a
discernible structure in spacetime? We shall discuss this intriguing aspect
of our theory in Sect.V . Presently, we turn to the second fundamental
issue, namely the construction of a quantum theory of strings based on the
statistical classical formulation developed so far. In the next section we
will build this theory from the ``~ground~'' up, in a literal sense, by
interpreting the fluid of relativistic strings as a model for the {\it vacuum}
in a stochastic quantum theory of strings.

\section{Semi-classical limit of the quantum loop equation}
\subsection{The string functional}

The primary objective of this section is to show how the classical
equations (\ref{duediciasette}) emerge as the $\hbar\rightarrow 0$ limit of
the  ``~string wave equation~''
\begin{equation}
\left[
{\hbar^2\over 2}\left(\int_0^1 ds\,\sqrt{x^{\prime\,2}}\right)^{-1}
\int_0^1 ds\,\sqrt{x^{\prime\,2}}\,
{\delta^2\over \delta\sigma^{\mu\nu}(s)
\delta\sigma_{\mu\nu}(s)}-m^4\right]\Psi[C]=0\ ,\qquad m^2\equiv
1/2\pi\alpha'\ .
\label{duediciotto}
\end{equation}
Here, $\Psi[C]$ denotes
a {\it functional} of the spatial loop $C:x^\mu=x^\mu(s)$
describing the string, and $\delta/\delta\sigma^{\mu\nu}(s)$,
denotes the functional derivative with respect to the area element.
Eq.(\ref{duediciotto}) is nothing but the string momentum mass shell
condition (\ref{media})
expressed through the quantum operators canonically conjugated
to $P_\mu(s)$ and $\Pi_{\mu\nu}(s)$ through the Correspondence Principle:
\begin{eqnarray}
&&P_\mu(s)\rightarrow i\hbar{1\over\sqrt{x^{\prime\,2}} }
{\delta\over\delta x^\mu(s)}\ ,\\
&&\Pi_{\mu\nu}(s)\rightarrow
i\hbar{\delta\over\delta\sigma^{\mu\nu}(s)}\ ,
\end{eqnarray}
and
\begin{equation}
{\delta\over\delta x^\mu(s)}=x^{\prime\,\nu}
{\delta\over\delta\sigma^{\mu\nu}(s)}\ .
\end{equation}

To simplify the notation, one can choose as loop parameter, $s$, the proper
length defined as $\displaystyle{ds^2=dx^\mu dx_\mu}$, so that
$\displaystyle{x^{\prime\,2}=1}$. Then, eq.(\ref{duediciotto}) takes the
form given by Hosotani \cite{yh}.

Before going into technical details, it may be useful to clarify in which
sense we ``~quantize~'' the string. As a matter of fact, when dealing with
extended
objects, there are at least two conceptually different ways to interpret
the meaning of the term quantization. The first and most used approach to
quantization treats the string as a mechanical system, quantizing its small
oscillations by canonical or path integral methods and interpreting the
resulting excitation
states of the string as particles of different mass and spin. Parallel to
this approach, there is the much less investigated geometrical framework,
which we wish to
explore here, where the string is not merely a mechanical device to generate
a particle spectrum,
but is regarded as a physical object in itself whose shape can quantum
mechanically fluctuate
between different configurations. From this viewpoint the wave equation
(\ref{duediciotto}) is the string counterpart of the Wheeler-DeWitt
equation,
in the sense that it determines the quantum geometry of the object
under investigation,
i.e. the string {\it spatial configuration} in our case, and the {\it spatial
metric} in the canonical
formulation of quantum gravity.
{\it From our own vantage point, to quantize the string means to determine
the
probability amplitude for the  string to
have a {\it spatial shape} described by the loop $C$.} At this point,
two objections to this interpretation can be immediately raised: first, the
functional $\Psi[C]$
is defined in an abstract loop-space, while one would like to define
a probability amplitude, or density, to find a string with a given shape
in the physical spacetime; second, even if we succeed in connecting
quantities defined in loop-space to their spacetime counterpart,
we expect to be unable to introduce a positive definite probability density
because of the relativistic invariance of the system.
While the first problem can be solved, at least formally, as we shall
see in the rest of this section, the second difficulty finds an acceptable
solution
only in a ``~second quantization~'' scheme, in which $\Psi[C]$
plays the role of field operator, creating and destroying loops of
a given shape. We shall come back to this
point at the end of this section. Meanwhile we shall still refer to
$\Psi[C]$ as  the ``~probability amplitude~'' to underscore the fact that we
are dealing with a first quantized theory in the semi--classical limit. We
believe that this approach not only is a useful introduction to
the second quantization of the string functional, but is also a necessary
intermediate step to bridge the gap between the quantum equation
(\ref{duediciotto}) and the classical equations (\ref{duequattordici}),
(\ref{duediciasette}). In this connection, we should mention Hosotani's {\it
loop variable} ansatz\cite{yh}
\begin{equation}
\Psi[C]\sim \exp\left({i\over\hbar}\oint_C A_\mu(x)dx^\mu\right)\ .
\label{hoso}
\end{equation}
This wave functional does serve the purpose of relating equation
(\ref{duediciotto}) with the classical
Hamilton--Jacobi\ equations for the string, but corresponds to a ``string plane
wave'' , i.e. to an unlocalized state in loop space that assigns an
equal probability amplitude to any loop shape. In the stochastic quantum
theory of strings that we have in mind, we are more interested
in a new family of states for which different geometric configurations are
weighted by a non trivial probability amplitude. This is by no means a
settled issue and therefore, without pretence of mathematical rigor, we
proceed simply by analogy with ordinary quantum mechanics, bearing in
mind that the real test of consistency of our approach is to derive the
classical Hamilton-Jacobi equations for the string starting from the wave
equation for the string functional. To this end, and in order to assign  a
probabilistic interpretation to $\Psi[C]$, which has
canonical dimensions of a $(length)^2$, we proceed in two steps: first we
must introduce
a suitable length scale in the normalization condition for $|\Psi[C]|^2$.
At this stage the only length scale we have at our disposal is provided
by the string tension $m^2=1/2\pi\alpha'$, and therefore we define
\begin{equation}
m^4\int D[C]\,\vert\Psi[C]\vert^2=1.
\label{norm}
\end{equation}
Next, we define the probability
density (~in Minkowski spacetime~) by inserting the identity
\begin{equation}
\int d^4x\,\delta\left[x-x(s)\right]=1\ ,
\label{ident}
\end{equation}
into the l.h.s. of eq.(\ref{norm}):
\begin{equation}
\int d^4x\, \overline{\rho}(x)=1\,
\rightarrow \overline{\rho}(x)\equiv
m^4 \int D[C]\,\delta^{4)}\left[x-x(s)\right]\,
\vert\Psi[C]\vert^2\ .
\label{dens}
\end{equation}
According to this definition, the function $\overline{\rho}(x)$ represents the
(probability) density of loops of arbitrary shape within an elementary
volume
element centered at the point $x$. Each different shape is weighted by the
factor $\vert\Psi[C]\vert^2$ and equation (\ref{dens})
provides the correct unit normalization for the total probability
density. With hindsight, we have denoted the quantum density by the same
symbol used for its classical counterpart. The reason for that will become
clear shortly.

The next step in our stochastic approach, which we describe in the next
subsection, is to show that the same probability density enters in the
semi-classical form of the regularized probability current derived from the
WKB string functional
\begin{equation}
\Psi[C]\equiv A[C]\exp
{i\over\hbar}\oint_C dy^\mu A_\mu(y)\ .
\label{dueventidue}
\end{equation}
This wave functional represents our own ansatz for a loop variable
describing a string state of (~area~) momentum
$\Pi_{\mu\nu}(=\partial_{[\mu}A_{\nu]})$ and amplitude $A[C]$, which we
assume to be slowly varying
over the loop space.

\subsection{Point-splitting regularization}

The stochastic quantum mechanical approach that we are setting up would
be ill defined without solving a technical problem of regularization related
to the non-local character of the loop field
(\ref{dueventidue}). Presently, our purpose is to suggest a possible
computational procedure to deal with the differential operator
in eq.(\ref{duediciotto}). In fact, a naive attempt to evaluate the second
functional derivative of the WKB ansatz (\ref{dueventidue}) results in a
term proportional to $\delta(0)$. This divergence can be kept under control
by regularizing eq.(\ref{duediciotto}) through point-splitting:
\begin{equation}
\lim_{\epsilon\rightarrow 0}\left[
{\hbar^2\over 2}\int_0^1 ds\, {\delta^2\over \delta\sigma^{\mu\nu}(s-
\epsilon/2)
\delta\sigma_{\mu\nu}(s+\epsilon/2)}-m^4\right]\Psi[C]=0\ ,
\label{dueventitre}
\end{equation}
where the limit $\epsilon\rightarrow 0$ is taken at the end of all
calculations. As long as
$\epsilon$ is non-vanishing, the two points $s_\pm\equiv s\pm
\epsilon/2$ can be considered as independent variables (~with respect to
$s$~) and the integration over the string parameter reduces to a
multiplication
by a constant equal to one.

Next, we multiply eq.(\ref{dueventitre}) by $\Psi^*[C]$ and (~functionally~)
integrate over the loop shape, then we split the real and imaginary
part of the equation thus obtained, and insert the WKB ansatz.
The limit $\epsilon\rightarrow 0$ is harmless in the real part of the
equation, and one finds
\begin{equation}
\int D[C]\int_0^1 ds A[C]\left[{\hbar^2\over 2}
{\delta^2 A[C]\over \delta\sigma^{\mu\nu}(s)\delta\sigma_{\mu\nu}(s)}
-A[C]\left({1\over 2}F_{\mu\nu}\left(x(s)\right)F^{\mu\nu}\left(x(s)\right)
+m^4\right) \right]=0\ .
\label{dueventiquattro}
\end{equation}
By inserting the unity (\ref{ident}) into (\ref{dueventiquattro}),
we can exchange
$F_{\mu\nu}\left(x(s)\right)$ for $F_{\mu\nu}(x)$. Then, equation
(\ref{dueventiquattro}) is satisfied at any spacetime and loop space
point whenever the quantity inside the square bracket is vanishing, i.e.
\begin{equation}
-{1\over 2}F_{\mu\nu}(x)F^{\mu\nu}(x)-m^4+{\hbar^2\over 2}\int_0^1 ds\,
{1\over A[C]}
{\delta^2 A[C]\over \delta\sigma^{\mu\nu}(s)\delta\sigma_{\mu\nu}(s)}=0\
,
\label{dueventicinque}
\end{equation}
which is nothing but equation (\ref{duediciasette})
plus quantum corrections of order $\hbar^2$. To recover the equation for the
divergence of the Gibbs current,
we note that in the limit of vanishing $\epsilon$:
\begin{equation}
Im\left\{\int_0^1ds\,\Psi^*[C]{\delta \over \delta\sigma^{\mu\nu}(s_-)}
{\delta\Psi[C] \over \delta\sigma_{\mu\nu}(s_+)}\right\}\sim
Im\left\{ \int_0^1ds\,{\delta \over \delta\sigma^{\mu\nu}(s_-)}\Psi^*[C]
{\delta\Psi[C] \over \delta\sigma_{\mu\nu}(s_+)}\right\}\ ,
\label{dueventisei}
\end{equation}
because the two terms containing first order derivatives cancel against
 each other. Taking now the limit $\epsilon\rightarrow 0$, we obtain the
``~continuity equation~'' for the current density in loop space:
\begin{equation}
{1\over 2}\int_0^1ds\,{\delta \over \delta\sigma^{\mu\nu}(s)}
J^{\mu\nu}\left[C;s\right]=0
\ ,\quad
J^{\mu\nu}\left[C;s\right]\equiv {\hbar\over 2i}\Psi^*[C]
{{\stackrel{\longleftrightarrow}{\delta}}\over
\delta\sigma_{\mu\nu}(s)}\Psi[C]\ .\label{dueventisette}
\end{equation}
This regularization procedure, artificial as it may seem, not only eliminates
the offending $\delta(0)$--term, but also leads to the
correct classical equations of motion. Indeed, inserting
eq.(\ref{dueventidue}) into eq.(\ref{dueventisette}) we obtain
\begin{equation}
J^{\mu\nu}\left[C;s\right]= A^2[C]\,F^{\mu\nu}\left(x(s)\right)\ ,
\label{dueventiotto}
\end{equation}
which is the WKB form of the loop space current density. The corresponding
spacetime current density is obtained as before by integrating eq.
(\ref{dueventiotto}) over all possible loop configurations through
the point $x$
and over the parameter $s$:
\begin{equation}
J^{\mu\nu}(x)\equiv
\int D[C]\,\delta^{4)}\left[x-x(s)\right]\,J^{\mu\nu}\left[C;s\right].
\end{equation}
Inserting again our WKB functional:
\begin{eqnarray}
J^{\mu\nu}(x)&=&
\int D[C]\,\delta^{4)}
\left[x-x(s)\right]\, A^2[C]\, F^{\mu\nu}(x)\nonumber\\
&=&\frac{\overline{\rho}(x)}{m^4}F^{\mu\nu}(x)\ .
\label{dueventinove}
\end{eqnarray}
The last equality in (\ref{dueventinove}) connects
the WKB current
density to the classical current through the identification of the
(quantum) density
$\overline{\rho}(x)$, as defined in
(\ref{dens}), with the classical density
$\overline{\rho}(x)=m^2\sqrt{-\gamma}\,Z(x)\,\mu(x)$.
Furthermore, if we require that the divergence of $J^{\mu\nu}(x)$
vanishes, then we obtain the classical equation (\ref{duediciasette}).
The general form of the quantum current in Minkowski spacetime can be
obtained
again summing over all the possible loop shapes
through the point $x$. The final result is the following correspondence
between classical and quantum quantities
\begin{equation}
\begin{array}{lcl}
J^{\mu\nu}(x)=
\int D[C]\,\delta^{4)}\left[x-x(s)\right]\,J^{\mu\nu}[C;s]
&\qquad{\buildrel\hbar\rightarrow 0\over\longrightarrow}\qquad &
W^{\mu\nu}(x)=(\overline{\rho}(x)/m^4)\, F^{\mu\nu}(x)\nonumber\\
\overline{\rho}(x)=
m^4\int D[C]\,\delta^{4)}\left[x-x(s)\right]\,|\Psi[C]|^2
&\qquad{\buildrel\hbar\rightarrow 0\over\longrightarrow}\qquad&
\overline{\rho}(x)=
m^2\sqrt{-\gamma}\,Z(x)\,\mu(x)
\nonumber\\
\partial_\mu J^{\mu\nu}(x)=0
&\qquad{\buildrel\hbar\rightarrow 0\over\longrightarrow}\qquad &
\partial_\mu \left(
\overline{\rho}(x) F^{\mu\nu}(x)\right)=0
\end{array}
\end{equation}

In the above code of correspondence we have imposed the vanishing of the
divergence of the current having in mind the analogy with the
continuity equation
in ordinary quantum mechanics. However,
when dealing with a relativistic system, such
a motivation is not justified.
In our case $\overline{\rho}(x)$ does not correspond to any of the
components of $J^{\mu\nu}(x)$; moreover,
as in the case of the Klein-Gordon equation for the pointlike particle,
a ``~probability~'' current such as $J^{\mu\nu}(x)$ is not positive
definite, and the assumption of being divergence free has no clear
motivation except that it allows to establish a relationship between
$J^{\mu\nu}(x)$ and
$W^{\mu\nu}$. At first glance, this may appear as an unsatisfactory feature
of the
model. Instead, we argue that this difficulty is merely the signal that the
quantum current (\ref{dueventisette}) is related to an underlying
gauge symmetry of a fully fledged quantum field theory of strings. The
natural candidate seems to be the generalized
gauge symmetry \cite{kr}
\begin{eqnarray}
\Psi'[C]&=&\Psi[C]\exp\left({i\over\hbar}\oint_C
\Lambda_\mu(x)dx^\mu\right)
\nonumber\\
\Psi^*{}'[C]&=&\Psi^*[C]
\exp\left(-{i\over\hbar}\oint_C \Lambda_\mu(x)dx^\mu\right)
\nonumber\\
A'_{\mu\nu}(x)&=&A_{\mu\nu}(x)-{1\over
g}\partial_{\,[\mu}\Lambda_{\nu]}\ .
\label{krg}
\end{eqnarray}
Then, the vanishing of $\partial_\mu J^{\mu\nu}(x)$ follows from
the transversality of $A_{\mu\nu}$. The symmetry (\ref{krg}) allows
to choose the {\it Lorenz gauge} $\partial^\mu A_{\mu\nu}=0$ to
dispose of the unphysical components of the Kalb-Ramond potential
and the generalized Maxwell equation requires the two-index current
to be divergence free:
\begin{equation}
\partial_\mu H^{\mu\nu\rho}(x)-g^2\overline{\rho}^2(x)A^{\nu\rho}(x)
=g J^{\nu\rho}(x)\ ,\rightarrow \partial_\mu J^{\mu\nu}(x)=0\ .
\end{equation}
   However, setting this problem aside, we conclude that the self-
consistency
of our semi-classical stochastic model requires a gauge invariant coupling
to a two-form potential $A_{\mu\nu}(x)$, and the interpretation
of the loop-functional as a field operator. Remarkably, the classical limit of
this loop field theory exists and corresponds to the {\it statistical}
Hamilton--Jacobi--Carath\'eodorytheory of strings.
\section{Stochastic Pregeometry}

In this section we wish to take one final step
towards
the construction of a
stochastic approach to the theory of relativistic strings. So far the theory
has been formulated in {\it Minkowski space} and this formulation may well
be adequate for particle physics. However, since strings may play an
important role in the problem of formation of structure in the universe, or
in the very early universe at superunification time, the assumption of a
featureless, preexisting spacetime continuum untouched by the dynamical
events unfolding in the universe, seems unwarranted. Therefore, our next
step is to relax that assumption. The point of view advocated here is that
the geometry of spacetime should not be assigned a priori, but should be
compatible, as in the case of gravity, with the matter content of the
universe. Thus, our specific objective is to establish: i) to what extent the
{\it dynamics} of a string network, interpreted as a stochastic
process\,
affects the geometry of spacetime and, conversely ii) how the structure of
spacetime affects the distribution of matter in the universe.
Interestingly enough, it turns out that the string dynamics as well as the
geometry of spacetime are determined by the same averaged least action
principle discussed in the previous sections. A result similar to ours was
obtained for {\it point-like particles} by Santamato \cite{san}, and our work
is a direct
extrapolation of that approach to the case of relativistic strings.
In a conventional approach to this problem, one would obtain the background
geometry of spacetime by integrating away the strings degrees of freedom
in the string functional\cite{frad}.
As an alternative, here we suggest a non minimal
coupling between the statistical string ensemble and the curvature of
spacetime. The paradigm of this approach is the generalized Nambu-Goto
action
\begin{eqnarray}
	I&=&-\int_D d^2{\xi} \left[m^4+\kappa R(x)\right]^{1/2}
	\sqrt{-\gamma}-\oint_C dx^\mu S\partial_\mu U\ ,\nonumber\\
	&=&-\int_D d^2{\xi} \left[\left(m^4+\kappa R(x)\right)^{1/2}
	\sqrt{-\gamma}-{1\over 2}
{\dot x}^{\mu\nu}\partial_{\,[\mu} S\partial_{\nu]} U\right]\ ,\nonumber\\
	 m^2&\equiv& 1/2\pi\alpha'\ ,\quad[\kappa]=(length)^{-2}\ .
\label{az}
\end{eqnarray}

In four dimensions, the {\it minimal} departure from the assumption of a
Riemannian geometry is the weaker assumption that spacetime is a generic
manifold with torsion free connections $\Gamma^\lambda{}_{\mu\nu}$ so
that the action (5.1) describes the coupling of the string degrees of freedom
to both metric and connection through the curvature scalar
$R(x)=g^{\mu\nu}(x)R_{\mu\nu}(\Gamma)$. The second term in (\ref{az}) is
the
now familiar boundary term introduced by
Carath\'eodory's formulation of the least
action principle. In this case, the complete Carath\'eodory figure is
defined by the Hamilton-Jacobi equations
       \begin{equation}
        {{\dot x}_{\mu\nu}\over\sqrt{-\gamma} }
        ={\partial_{\,[\mu} S\partial_{\nu]} U\over\sqrt{m^4+\kappa R}}\ ,
	\end{equation}
	\begin{equation}
        {1\over 2}\partial_{\,[\mu} S\partial_{\nu]} U
	\partial^{\,[\mu} S\partial^{\nu]} U =
        -\left(m^4+\kappa R\right)\ .\label{hjr}
        \end{equation}

The solutions of the first equation represent the family of extremal
surfaces while the second equation governs the evolution of the wave fronts
$S(x)=const.$, $U(x)=const.$ The passage to a statistical string ensemble
follows the same steps described in the previous sections. Since most of
the details are the same, we outline the main steps:

1) consider a two--parameter family of world-sheets
$x^\mu=x^\mu({\xi}^a;\theta^i)$, $i=1,2$
randomly distributed in parameter space according to a given distribution
function $\mu=\mu(\theta^i)$.

2) Averaging over the above parameters, one
obtains the action for the statistical string ensemble
	\begin{equation}
	\langle I \rangle=-\int_D d^2{\xi} \int d^2\theta\, \mu(\theta^i)\,
	\left[\left(m^4+\kappa R\right)^{1/2}
	\sqrt{-\gamma}-{1\over 2}
	{\dot x}^{\mu\nu}\partial_{\,[\mu} S\partial_{\nu]} U\right]
	\end{equation}

3) The action is now defined as an integral over
four-dimensional {\it spacetime} by using the following code
of correspondence and transformation rules
	\begin{eqnarray}
	&&x^\mu\leftrightarrow \tau,\sigma,\theta^1,\theta^2\ ,
	\quad \mu(x)=\mu(\theta^i(x))\ ,\nonumber\\
	&&d^2{\xi}\, d^2\theta\,\mu(\theta^i)= d^4x\,Z(x)\,\mu(x)\ ,
	\qquad Z(x)\equiv
	\frac{\partial(\tau,\sigma,\theta^1,
	\theta^2)}{\partial(x^0, x^1, x^2, x^3)}\ ,\nonumber\\
	&&\rho(x)\equiv Z(x)\,\mu(x)\ ,\quad
	W^{\mu\nu}(x)\equiv \rho(x)\,{\dot x}^{\mu\nu}(x)\ .
	\end{eqnarray}

The result of this procedure is the following action defined solely in terms
of field variables
        \begin{eqnarray}
	\langle I \rangle &=&-\int_D d^4x
	\left[\left(m^4+\kappa R\right)^{1/2}
	\sqrt{-{1\over 2}W^{\mu\nu}W_{\mu\nu}}-{1\over 2}
W^{\mu\nu}\partial_{\,[\mu} S\partial_{\nu]} U\right]\ ,\nonumber\\
	&=&-\int_D d^4x
	\left[\left(m^4+\kappa R\right)^{1/2}
	\sqrt{-{1\over 2}W^{\mu\nu}W_{\mu\nu}}-{1\over 2}
	W^{\mu\nu}\nabla_{\,[\mu} A_{\nu]}\right]\ .
	\end{eqnarray}

4) Varying this action with respect to $W^{\mu\nu}$
and
$A_\nu$ yields the field equations

        \begin{equation}
	\nabla_\mu W^{\mu\nu}=0\ ,
	\end{equation}
	\begin{equation}
	\left(m^4+\kappa R\right)^{1/2}{W_{\mu\nu}\over\sqrt{-|W|^2}}
	=\nabla_{\,[\mu} A_{\nu]}\ .
	\end{equation}

Since we have assumed a torsion-free connection, we can replace the
covariant derivatives in $F_{\mu\nu}$ above with ordinary derivatives.
Then, if the functions $S(x)$ and $U(x)$ satisfy the Hamilton-Jacobi
equation (\ref{hjr})
that is,
\begin{equation}
F_{\mu\nu}F^{\mu\nu}=-2(m^4+\kappa R)\ ,
\end{equation}
then $W^{\mu\nu}$
can be identified with the {\it Gibbs current density} associated with the
statistical ensemble
	\begin{equation}
 W^{\mu\nu}={1\over m^4}\overline{\rho}(x)\,\sqrt{-g}\,g^{\mu\rho}g^{\nu\sigma}
	\,F_{\rho\sigma}\ .
	\end{equation}

With this expression for the current density, the action for the string
network takes the remarkably simple form

        \begin{equation}
\langle I \rangle_{cl.}=-{1\over m^4}\int d^4x\,\sqrt{-g}\,\overline{\rho}(x)
	\left[m^4+\kappa R(x)\right]
	\end{equation}

This result is precisely analogous to that obtained by
Santamato\cite{san}
in the case
of a statistical ensemble of relativistic point-like particles. Therefore, we
can draw a similar conclusion: variation of the above action with respect to
the connection yields
        \begin{equation}
	\delta_\Gamma  \langle I \rangle_{cl.}=0\ ,\rightarrow
 	\Gamma^\lambda{}_{\mu\nu}=\{_\mu{}^\lambda{}_\nu\}+{1\over 2}
	\left(\phi_\mu\delta^\lambda_\nu+\phi_\nu\delta^\lambda_\mu-
	g_{\mu\nu}g^{\lambda\rho}\phi_\rho \right)\
	\end{equation}
which represents the anticipated deviation from the Riemannian geometry in
the form of a Weyl-Christoffel connection. In all fairness and with
hindsight, this result is hardly surprising since it follows whenever a first
order, or Palatini method is employed in any model where the scalar
curvature
is directly coupled to some matter field. The novelty of our result is
that the $\overline{\rho}(x)$ field, rather than being a
generic field variable in a
scalar--tensor theory of gravity, represents the macroscopic string
energy density.
With this specific interpretation of the $\overline{\rho}(x)$ field in mind,
there are two
aspects of the result (5.12) which are worth emphasizing. First, the {\it
Weyl
field} $\phi_\mu(x)$ \ is longitudinal, or pure gauge, since
        \begin{equation}
	\phi_\mu(x)=
\overline{\rho}{}^{\,-1}\partial_\mu
\overline{\rho}=\partial_\mu\ln(
\overline{\rho}/M^4)\ ,
	\end{equation}
where $M$ represents an arbitrary constant. This is not an assumption but
a dynamical consequence of the theory. Therefore the length of a vector
remains
constant after parallel displacement along a closed path
        \begin{equation}
	L^2=L_0^2\exp\left(\oint_\gamma dx^\mu \phi_\mu(x)\right)=L^2_0\
,
	\end{equation}
thereby removing, at least in this theory, one of the main criticisms about
the physical relevance of Weyl's geometry.
The second and most important point is that this newfound geometry does
not extend over the entire spacetime; rather, it is confined to those regions
where the string density fluctuates or changes appreciably. In contrast,
wherever
           $\displaystyle{
\overline{\rho}(x)\simeq const.\equiv
\overline{\rho}_0\rightarrow
	\phi_\mu\simeq 0}$
and our classical action reduces to Einstein's action with a cosmological
constant

          \begin{equation}
	\langle I \rangle_{cl.}=-{1\over m^4}\int d^4x\,\sqrt{-g}\,
\overline{\rho}_0\, \left[m^4+\kappa R(g)\right]\ ,
	\end{equation}
if we identify $R(g)$ with Ricci's {\it Riemannian} scalar curvature
and
         \begin{equation}
	{\kappa\overline{\rho}_0\over m^4}\equiv {1\over 16\pi G_N}\ ,\qquad
	\overline{\rho}_0=-{\Lambda\over 8\pi G_N}\ .
	\end{equation}
	We would like to conclude this section, and the paper, with few
speculative remarks on the possible consequences of the above results on
the longstanding
problem of formation of structure in the universe. According to our
discussion, spacetime  seems to have
a multiphase, or {\it cellular} structure such that regions in a Riemannian
geometric phase in which the string density is roughly constant, are
connected by
Weyl regions, or {\it domain walls} over which the string density changes
abruptly. The Weyl geometry appears to be the geometry of the {\it domain
walls} between Riemannian cells of spacetime. One possible picture that
this structure brings to mind is the observed{\it soap--bubble--like}
pattern of cosmic voids with vanishing energy density, separated by domain
walls on
which
filamentary, or string-like superclusters of galaxies seem to be
concentrated. Admittedly, this is a purely qualitative picture. However, it
emerges from a completely analytical approach to the dynamics of a string
network. Of course, an in--depth analysis is needed, possibly using computer
simulations, in order to corroborate its validity.

Another possible scenario that the cellular structure of spacetime brings to
mind, is that of {\it chaotic inflation}. Following Linde \cite{linde}, in this
case we envisage the whole Universe as a cluster of microuniverses, some
of them inflationary depending on the value and sign of the string energy
density as given by equation (5.16). A typical initial size of a randomly
chosen spacetime cell should be of the order of Planck's length with an
energy density of the order of Planck's density. The classical evolution of
any such cell has been analyzed in detail in an earlier paper \cite{aps}. The
present paper goes one step further, at least at a conceptual level: the
controlling factor here is the stochastic field
$\overline{\rho}(x)$ which appears in the  expression of
Weyl's field and represents the string density in spacetime. We cannot
overemphasize the fact that this stochastic field originates at the quantum
level and
represents the classical counterpart of the probability density $|\Psi[C]|^2$
in loop space. Thus, the {\it dynamically induced} Weyl-Riemann geometry of
spacetime
that
we have uncovered, should be regarded as the result of an averaging process
over the string quantum fluctuations and, ultimately, it may originate from
the stochastic nature of spacetime itself. In this sense, the present work
may well provide a quantum basis for stochastic inflation. Finally, to the
extent that the
classical limit of the Wheeler-DeWitt string wave equation exists and that
this limit represents a {\it statistical} theory, we claim that we have
developed also a
{\it stochastic} quantum string theory.

\end{document}